\documentstyle[aps,prc,bm]{revtex}

\newcommand{\N}{N_{\rm evts}}
\newcommand{\Np}{N'}
\newcommand{\mean}[1]{\left\langle #1 \right\rangle}

\begin{document}

\preprint{ULB-TH-01/028}
\draft

\title{Flow analysis from cumulants: a practical guide}

\author{\underline{Nicolas Borghini},$^{1}$
Phuong Mai Dinh,$^2$ 
and Jean-Yves Ollitrault$^{2,3}$}

\address{$^1$~Service de Physique Th\'eorique, CP225, 
Universit\'e Libre de Bruxelles, 1050 Brussels, Belgium}
\address{$^2$~Service de Physique Th\'eorique, CEA-Saclay, 
91191 Gif-sur-Yvette cedex, France}
\address{$^3$~L.P.N.H.E., Universit\'e Pierre et Marie Curie, 
4 place Jussieu, 75252 Paris cedex 05, France}

\maketitle

\begin{abstract}
We have recently proposed a new method of flow analysis, based on a cumulant 
expansion of multiparticle azimuthal correlations. 
Here, we describe the practical implementation of the method.
The major improvement over traditional methods is that the cumulant expansion 
eliminates order by order correlations not due to flow, which are often large 
but usually neglected.
\end{abstract}


\section{Introduction}
\label{s:intro}

The measurement of the azimuthal distributions of outgoing particles with 
respect to the reaction plane  in noncentral heavy ion 
collisions---the flow analysis---is an important probe of the
interaction region of the collision \cite{flow-reviews}.
In particular, it has raised much interest at ultrarelativistic 
energies \cite{Poskanzer} where it may signal the formation 
of a quark-gluon plasma \cite{QGPfromFlow}. 
In addition, combining flow and two-particle interferometry results yields a 
three-dimensional picture of the emitting source \cite{Flow&HBT,Retiere}. 
Therefore, accurate flow measurements are highly needed. 

Azimuthal distributions are characterized by the Fourier 
coefficients \cite{Voloshin:1996mz}
\begin{equation}
\label{def_vn}
v_n \equiv \mean{e^{in(\phi-\Phi_R)}} = \mean{\cos n(\phi-\Phi_R)}, 
\end{equation}
where $\phi$ is the azimuthal angle of an emitted particle in the laboratory 
frame, $\Phi_R$ is the azimuth of the reaction plane, and angular brackets 
denote a statistical average over many particles and events. 
Ideally, $v_n$ should be measured for various particles 
as a function of their transverse momentum $p_T$ and rapidity $y$
(``differential'' flow).  
The first two harmonics $v_1$ and $v_2$ are the so-called directed 
and elliptic flows, respectively.

Since the azimuth $\Phi_R$ in a given event is unknown, the coefficients $v_n$ 
are extracted from the azimuthal correlations between outgoing particles. 
The underlying idea is that the correlation of every particle with the reaction 
plane induces correlations between the particles. 
Standard methods extract flow from two-particle azimuthal correlations, either 
directly \cite{Correlation-f}, or through the correlation between two 
``subevents'' \cite{subevents,standard}.
However, the correlation between two given particles is not only due to flow, 
and the other sources of correlation---as, e.g., quantum Bose-Einstein 
effects, momentum conservation, resonance decays, jets---may dominate 
the measured signal, especially for peripheral collisions, thereby spoiling the 
validity of the analysis \cite{Dinh:2000mn,Borghini:2000cm}. 
The impact of ``nonflow'' correlations on the flow analysis might tentatively 
be minimized: cuts in phase space can be used to avoid the influence of quantum 
effects and resonance decays, while the contribution of momentum conservation 
to the measured correlation can be calculated and subtracted 
\cite{Borghini:2000cm,Danielewicz:1988in}. 
However, the various recipes require some {\em a priori} knowledge of nonflow 
correlations; furthermore it is necessary to assume that {\em all} sources of 
such correlations are known and accounted for, which may not be true. 

To remedy the contamination from nonflow correlations, which amounts to 
systematic uncertainties on the flow values, we have introduced new methods of 
flow analysis, based on a cumulant expansion of multiparticle azimuthal 
correlations \cite{Borghini:2001sa,Borghini:2001vi}. 
The principle of the methods is that when cumulants of higher order are 
considered, the relative contribution of nonflow effects, and thus the 
corresponding systematic error,  decreases. 
Recently, this cumulant expansion has been successfully applied by 
the STAR Collaboration at RHIC \cite{Poskanzer,Tang:2001yq}. 

More precisely, the cumulant of $2k$-particle azimuthal correlations, 
which we denote $c_n\{2k\}$ (where $n$ is the Fourier harmonic and 
$2k$ is an even integer, in practice 2, 4 or 6), 
is a quantity built with all the measured azimuthal correlations up 
to order $2k$, i.e., the 
$\mean{\exp [in(\phi_1+\phi_2+\cdots+
\phi_{k'}-\phi_{k'+1}-\cdots-\phi_{k'+k''})]}$, with $k'+k''\leq 2k$.  
The key feature of the cumulant is that it eliminates the contribution 
of lower order correlations, so that only the genuine
$2k$-particle correlation remains. 
Flow, which is essentially a collective effect, gives a contribution 
to the cumulant proportional to $v_n^{2k}$. 
The remaining contribution, from $2k$-particle nonflow correlations, 
scales as $N^{1-2k}$, 
where $N$ is the total multiplicity of particles emitted in an event 
\cite{Borghini:2001sa}. 
Therefore, the flow dominates if 
\begin{equation}
\label{systematic}
v_n^{2k} \gg \frac{1}{N^{2k-1}} \qquad \Leftrightarrow \qquad 
v_n \gg \frac{1}{N^{1-1/2k}}. 
\end{equation}
Stated differently, if the cumulant $c_n\{2k\}$ is much larger than 
$N^{1-2k}$, 
it is dominated by flow, and therefore yields an estimate of $v_n$, which we 
denote $v_n\{2k\}$, with a systematic error 
(due to unknown nonflow correlations)  
of order ${\cal O}((Nv_n)^{1-2k})$. 
When $k$ increases, the systematic error decreases: this is why the estimate 
$v_n\{4\}$ derived from the fourth cumulant is {\em a priori} more accurate 
than $v_n\{2\}$, which is in fact the value given by two-particle methods. 

In the following, we describe the practical implementation of the method, 
referring the reader to Refs.~\cite{Borghini:2001sa,Borghini:2001vi} 
for further theoretical justifications. 
The first step 
consists in deriving a global measurement of $v_n$, integrated over some 
phase-space region, typically a detector acceptance, for a given centrality 
class (Sec.~\ref{s:int}):
this is equivalent to 
reconstructing the reaction plane and  obtaining 
the ``event plane resolution'' in the subevent method \cite{subevents}. 
This ``integrated'' flow serves as reference for the 
differential flow analysis discussed in Sec.~\ref{s:diff}.
An important feature of our method is 
that it automatically takes into account azimuthal 
inhomogeneities in the acceptance of the detector. 
In the case of a detector with only partial azimuthal coverage, 
minor modifications occur, which are 
given in Sec.\ \ref{s:acceptance}.

\section{Integrated flow}
\label{s:int}

Consider a data set of $\N$ events of approximately the 
same centrality, recorded in a run with a constant detector
acceptance. 
In this section, we explain how estimates of the flow, 
integrated over the detector acceptance, can be obtained from 
cumulants of 2-, 4-, and 6-particle correlations. 

We denote by $\phi_j$ the azimuths of the outgoing particles with respect to a 
fixed direction in the laboratory.
The various quantities of interest are constructed from 
the real-valued generating function 
\cite{Borghini:2001vi}
\begin{equation}
\label{gen_func}
G_n(z) = \prod_{j=1}^M \left[ 1+\frac{w_j}{M} 
\left( z^* e^{in\phi_j} + z e^{-in\phi_j} \right) \right] = 
\prod_{j=1}^M \left[ 1+\frac{w_j}{M} 
\left( 2x\cos (n\phi_j) + 2y \sin(n\phi_j) \right) \right], 
\end{equation}
where the product runs over $M$ particles detected in a single event 
and $z=x+iy$ is an arbitrary complex number.   
This generating function has no physical meaning in itself, 
but after averaging over events, 
the coefficients of its expansion in powers of $z$ and $z^*\equiv x-iy$ 
yield multiparticle azimuthal correlations of arbitrary orders.
In practice, the variable $z$ corresponds to interpolation points 
used to estimate the 
various quantities encountered in the analysis, as will be explained shortly. 

As in the standard flow analysis, 
a weight $w_j$ is attributed to particle $j$, which is a function   
of particle type, transverse momentum, and rapidity. 
Naturally, the integrated flow obtained from this generating 
function will be weighted by $w$, i.e. 
$V_n\equiv\mean{w\,e^{in(\phi-\Phi_R)}}$. 
The weight  must be chosen so as to maximize the effects of flow relative to 
statistical fluctuations. 
As we shall see below, this is achieved by maximizing the dimensionless quantity 
$\chi_n\equiv V_n\sqrt{M/\mean{w^2}}$ (this quantity also characterizes the 
event plane resolution in the standard flow analysis \cite{standard}). 
As a consequence, the optimal weight for a given particle is its flow 
$v_n(p_T,y)$ itself \cite{Borghini:2001sa}.        
Thus, a thorough flow analysis should go twice through Secs.\ \ref{s:int} and 
\ref{s:diff}. 
The first time, integrated flow can be extracted using some reasonable guess 
for the weights, thereby obtaining values for $v_n(p_T,y)$; in turn, these 
values will serve as weights in the second, final analysis \cite{Chung:2001je}.

In Eq.~(\ref{gen_func}), the number $M$ of particles should be the same for all 
events: in each event, a set of $M$ particles must be randomly chosen out of 
the $M_{\rm tot}$ detected particles. 

In order to obtain the cumulants, one first averages $G_n(z)$ over events, 
which yields an average generating function $\mean{G_n(z)}$. 
We then define~\cite{Borghini:2001vi}  
\begin{equation}
\label{int_cum}
{\cal C}_n(z) \equiv M \left[ \mean{G_n(z)}^{1/M} - 1 \right].
\end{equation}
The cumulant of $2k$-particle correlations $c_n\{2k\}$ is the coefficient of 
$z^kz^{*k}/(k!)^2$ in the power-series expansion of ${\cal C}_n(z)$. 
To construct the first three cumulants, one may truncate the series to order 
$|z|^6$ and compute ${\cal C}_n(z)$ at the following interpolation points:   
\begin{equation}
\label{points}
z_{p,q} = x_{p,q} + iy_{p,q}, \qquad 
x_{p,q} \equiv r_0\sqrt{p}\,\cos\left(\frac{2q\pi}{q_{\rm max}}\right), \qquad
y_{p,q} \equiv r_0\sqrt{p}\,\sin\left(\frac{2q\pi}{q_{\rm max}}\right), 
\end{equation}
for $p=1,2,3$ and $q=0,\ldots, q_{\rm max}-1$, where $q_{\rm max}\ge 8$. 
The parameter $r_0$ must be chosen as a compromise between errors due to higher 
order terms in the power-series expansion, which rapidly increase with $r_0$, 
and numerical errors. 
Assuming that the numerical error is proportional to the total number of 
elementary operations performed (of order $M\N$), we obtain the estimate 
$r_0\simeq (\epsilon \N^{3/2}M)^{1/8}\sqrt{M/\mean{w^2}}$ where $\epsilon$ is 
the accuracy of elementary operations, typically $10^{-16}$ in double precision. 
This gives $r_0\sim 2$ with weights of order unity and standard values of the 
multiplicity ($M\sim 300$) and number of events ($\N\sim 20000$). 

Once the values ${\cal C}_n(z_{p,q})$ have been computed, they must be 
averaged over the phase of $z$: 
\begin{equation}
C_p \equiv 
\frac{1}{q_{\rm max}} \sum_{q=0}^{q_{\rm max}-1} {\cal C}_n(z_{p,q}), \qquad
p=1,2, 3. 
\end{equation}
The cumulants of 2-, 4- and 6-particle correlations are then given respectively 
by 
\begin{equation}
\label{cum_int}
c_n\{2\} = {1\over r_0^2} \left(3\, C_1-{3\over 2}C_2+{1\over 3}C_3\right), 
\qquad
c_n\{4\} = {2\over r_0^4} \left(-5\, C_1+4\,C_2-\,C_3\right), \qquad
c_n\{6\} = {6\over r_0^6} \left(3\, C_1-3\,C_2+C_3\right).
\end{equation}

These cumulants are related to the weighted integrated flow 
$V_n\equiv\mean{w e^{in(\phi-\Phi_R)}}$. 
This is the point where acceptance considerations come into play. 
If the detector has full azimuthal coverage, each cumulant $c_n\{2k\}$ gives 
an estimate of the corresponding $V_n$, which we denote by $V_n\{2 k\}$:
\begin{equation}
\label{c_and_v}
V_n\{2\}^2 = c_n\{2\}, \qquad 
V_n\{4\}^4 = -c_n\{4\}, \qquad
V_n\{6\}^6 = c_n\{6\} / 4. 
\end{equation}
Generalized relations valid for detectors with partial azimuthal coverage or 
efficiency are given in Sec.\ \ref{s:acceptance}. 

In practice, the use of higher order cumulants is often 
limited by statistics. The order of magnitude of statistical errors
can easily be estimated. 
The computation of the cumulant $c_n\{2k\}$ 
relies on the choice of $2k$ particles among $M$ in each of the $\N$ available 
events, i.e., it involves roughly $M^{2k}\N$ $(2k)$-uplets of particles. 
Taking into account the weights $w$, the resulting statistical uncertainty 
on $c_n\{2k\}$ is of order ${\cal O}(\mean{w^2}^k/\sqrt{M^{2k}\N})$. 
The relative error on $V_n\{2k\}$ is thus of order
\begin{equation}
{\delta V_n\{2k\}\over V_n\{2k\}}\sim {1\over \chi_n^{2k}\sqrt{\N}}.
\end{equation}
with $\chi_n \equiv V_n \sqrt{M/\mean{w^2}}$. 
A thorough calculation, performed in Ref.\ \cite{Borghini:2001vi}, Appendix~D, 
shows that this order of magnitude is indeed correct as long 
as $\chi_n$ is not larger than unity, which is the case for 
most experiments at ultrarelativistic energies.

In order to increase the statistics, it is possible to combine different runs
performed in a given experiment, each of which has its own characteristics 
(different orientation of the magnetic field, etc.).        
For each of these runs, following the procedure described above leads to 
cumulants $c_{n,\alpha}\{2k\}$ (where $\alpha$ labels the run) and finally,
accounting for the specific acceptance corrections, 
to flow estimates $V_{n, \alpha}\{2k\}^{2k}$. 
The proper way to combine the runs consists in averaging the 
$V_{n,\alpha}\{2k\}^{2k}$ ({\em not the} $V_{n,\alpha}\{2k\}$) 
of the various runs, weighted by the number of events in each run. 
Note that if statistical fluctuations are large, it may happen that a run 
yield a negative value of $V_{n,\alpha}\{2k\}^{2k}$. 
Such a run must nevertheless be included in the averaging procedure.

\section{Differential flow}
\label{s:diff}

Let us turn to differential flow, i.e., to the flow of an identified particle 
in a restricted portion of phase space (typically a narrow $p_T$ or $y$ 
interval). 
We call such a particle a ``proton'', and denote its azimuth by $\psi$; thus, 
the differential flow is $v'_n\equiv\mean{e^{in(\psi-\Phi_R)}}$, where the 
average value is taken over all protons. 

Experimentally, the differential flow is obtained from the azimuthal 
correlations between the proton and the particles previously used to determine 
the integrated flow, which we call ``pions''. 
It is well known in the standard flow analysis that from the reaction plane 
reconstructed in harmonic $n$, one may reconstruct not only the corresponding 
$v'_{n}$, but also higher harmonics $v'_{mn}$, where $m$ is an integer (in 
practice, $m=1$ or $2$).
Here, this is done by correlating the proton with $m$ pions, i.e., by measuring 
$\mean{\exp[in(m\psi-\phi_1-\cdots-\phi_m)]}$. 
As in the case of integrated flow, nonflow contributions to this correlation 
can be eliminated by going to higher orders, i.e., by correlating the proton 
with $2k+m$ pions with $k\ge 0$ (in practice $k=0$ or $1$), and constructing a 
cumulant $d_{mn/n}\{2k+m+1\}$. 
The subscript refers to the fact that $v'_{mn}$ is measured by using pions in 
harmonic $n$, while the number in curly brackets is the order of the 
correlation: $2k+m$ pions and 1 proton. 

This cumulant retains only the contributions from flow, proportional to 
$v'_{mn} V_n^{2k+m}$, and from ($2k+m+1$)-particle nonflow correlations, 
which scale like ${\cal O}(1/N^{2k+m})$: 
all lower order nonflow correlations have been removed. 
If the contribution of flow dominates, the cumulant $d_{mn/n}\{2k+m+1\}$ yields 
an estimate of $v'_{mn}$, which we naturally denote $v'_{mn/n}\{2k+m+1\}$. 

As in the case of integrated flow, we derive the cumulants $d_{mn/n}\{2k+m+1\}$ 
using a generating function, namely
\begin{equation}
\label{gen_func_diff}
{\cal D}_{mn/n}(z) \equiv \frac{\mean{e^{imn\psi}\,G_n(z)}}{\mean{G_n(z)}}. 
\end{equation}
The cumulant $d_{mn/n}\{2k+m+1\}$ is the real part of the coefficient of 
$z^{*k} z^{k+m} / [k! (k+m)!]$ in the power-series expansion of 
${\cal D}_{mn/n}(z)$. 
In the numerator of Eq.\ (\ref{gen_func_diff}), the average is performed over 
all protons (i.e., an event with 2 protons is counted twice; this was not stated 
correctly in \cite{Borghini:2001vi}). 
On the other hand, the denominator is averaged over all events. 
If the proton is one of the ``pions'', i.e., if it was used in the calculation 
of the generating function Eq.\ (\ref{gen_func}), one should divide $G_n(z)$ 
by $1+ w_j ( z^* e^{in\psi} + z e^{-in\psi})/M$, where $\psi$ is the proton 
azimuth, to avoid autocorrelations. 
Note that while the number of pions in Eq.\ (\ref{gen_func}) was fixed, the 
number of protons must be allowed to fluctuate from event to event: to 
increase statistics, one should use all available protons. 

To extract the cumulants, one computes the product $z^{*m}{\cal
D}_{mn/n}(z)$ at the points
$z_{p,q}$, Eq.\ (\ref{points}); one then takes the real part, 
and averages over angles: 
\begin{equation}
D_p\equiv \frac{\left (r_0\sqrt{p}\right)^m}{q_{\rm max}}
\sum_{q=0}^{q_{\rm max}-1}
\left[\cos\left(m\frac{2\,q\,\pi}{q_{\rm max}}\right) \, X_{p,q}+
\sin\left(m\frac{2\,q\,\pi}{q_{\rm max}}\right) \,Y_{p,q}\right],
\end{equation}
with $p=1,2, 3$ and $X_{p,q}+i Y_{p,q} \equiv {\cal D}_{mn/n}(z_{p,q})$.
Note that although we present the integrated and differential flow analyses as 
two successive steps, they are in fact simultaneous: while the generating 
function $G_n(z)$ is calculated for a given event, one can at the same time 
compute its product by $e^{imn\psi}$ for the numerator of 
Eq.\ (\ref{gen_func_diff}). 

For $m=1$ (useful for $v'_{1/1}$ or $v'_{2/2}$), the lowest order
cumulants are given by 
\begin{equation}
d_{n/n}\{2\}={1\over r_0^2}\left(2\, D_1-{1\over 2}\, D_2\right),\qquad
d_{n/n}\{4\}={1\over r_0^4}\left( -2\, D_1+ D_2\right),
\end{equation}
while for $m=2$, which can be used to derive $v'_{2/1}$,  
\begin{equation}
d_{2n/n}\{3\}={1\over r_0^4}\left(4\, D_1-{1\over 2}\, D_2\right),\qquad
d_{2n/n}\{5\}={1\over r_0^6}\left( -6\,D_1+{3\over 2}\, D_2\right). 
\end{equation}

These cumulants must then be related to the differential flow. 
For a perfect detector:
\begin{eqnarray}
\label{d_and_v'}
v'_{n/n}\{2\} = d_{n/n}\{2\} / V_n, &\qquad&
v'_{n/n}\{4\} = -d_{n/n}\{4\} / V_n^3,\cr
v'_{2n/n}\{3\} = d_{2n/n}\{3\} / V_n^2,&\qquad&
v'_{2n/n}\{5\} = -d_{2n/n}\{5\} / (2V_n^4).
\end{eqnarray}
Note that these relations involve the integrated flow $V_n$ obtained in the 
previous section. 
Relations for a nonisotropic acceptance are given in next Section. 
The order of magnitude of the statistical uncertainty can be estimated using 
the same arguments as for integrated flow. 
One obtains 
\begin{equation}
\delta v'_{mn/n}\{2k+m+1\}\sim {1\over\chi_n^{2k+m}\sqrt{\Np}},
\end{equation}
where $\Np$ is the number of protons used in the analysis.
It may be worth noticing that the statistical error does not only depends on 
the available event statistics and multiplicity. 
It also strongly depends on the flow itself through the parameter $\chi_n$.

\section{Acceptance corrections}
\label{s:acceptance}

Our method can be used even when the detector used to measure the particle 
azimuths has only partial azimuthal coverage, provided that the event 
centralities be determined with an {\em independent} detector, as, e.g., a ZDC, 
with an approximately isotropic coverage. 
This is to make sure that the apparent multiplicity/centrality is not biased by 
the orientation of the reaction plane with respect to the detector. 

A specific detector is characterized by its acceptance/efficiency function 
$A(j,\phi,p_T,y)$, which represents the probability that a particle of type $j$ 
(pion, proton, etc.) with azimuth $\phi$, transverse momentum $p_T$, and 
rapidity $y$, be detected. 
In practice, $A(j,\phi,p_T,y)$ is proportional to the number of hits in a 
$(\phi, p_T, y)$ bin, and thus can be obtained in a straightforward way while 
scanning through the data. 
The Fourier coefficients of the acceptance function are 
\begin{mathletters}
\label{def_ak}
\begin{equation}
\label{ak_pTy}
A_n(j,p_T,y) \equiv \displaystyle \int_0^{2\pi}\! A(j,\phi,p_T,y) \, 
e^{-in\phi} \, {\rm d}\phi. 
\end{equation}
These differential coefficients can be integrated, with appropriate weighting 
and a sum over the various types of particles used for the flow analysis, so as 
to describe the ``integrated'' acceptance of the detector: 
\begin{equation}
\label{ak}
a_n = \frac{\displaystyle \sum_j \int 
w(j,p_T,y)\,A_n(j,p_T,y)\, {\rm d}p_T{\rm d}y}{\displaystyle \sum_j
\int \, w(j,p_T,y)\,A_0(j,pT,y)\, {\rm d}p_T{\rm d}y}. 
\end{equation}
\end{mathletters}
Note our introducing the weights $w(j,p_T,y)$, which are of course the same as 
in Eq.\ (\ref{gen_func}). 

When the acceptance of a detector is not perfectly isotropic in $\phi$, the 
cumulants Eq.\ (\ref{cum_int}) will mix different flow harmonics: $c_n\{2k\}$ 
does not depend only on $V_n$, but also on other $V_p$ with $p \neq n$. 
For instance, 
\begin{mathletters}
\label{cum4}
\begin{eqnarray}
c_1\{4\} & = &  
- \left[(1-|a_1|^2)^4 + 4\,(1-|a_1|^2)^2 |a_2-a_1^2|^2 +|a_2-a_1^2|^4\right] 
V_1^4 \cr
 & & - \left[|a_1-a_2a_1^*|^4 + 4\,|a_1-a_2a_1^*|^2  |a_3-a_1a_2|^2 +
|a_3-a_1a_2|^4 \right] V_2^4, \label{cum41} 
\end{eqnarray}
and 

\begin{eqnarray}
c_2\{4\} & = &  
-\left[|a_1-a_2a_1^*|^4 + 4\,|a_1-a_2a_1^*|^2 |a_3-a_1a_2|^2 +
|a_3-a_1a_2|^4 \right] V_1^4 \cr
& & -\left[ (1-|a_2|^2)^4 + 4\,(1-|a_2|^2)^2  |a_4-a_2^2|^2 + |a_4-a_2^2|^4 
\right] V_2^4, \label{cum42}
\end{eqnarray}
\end{mathletters}
where we have assumed that all other flow harmonics $V_{n \geq 3}$ are 
negligible. 
The corresponding relations for the cumulants $c_1\{2\}$ and $c_2\{2\}$ can be 
found in Ref.\ \cite{Borghini:2001vi}, Appendix C1. 
If the detector acceptance is not too bad, the coefficients $a_{k\neq 0}$ will 
be small, and these expressions are close to Eqs.\ (\ref{c_and_v}), valid for 
perfect detector. 

Equations (\ref{cum4}) form a linear system which can easily be inverted to 
express $V_1^4$ and $V_2^4$ (or, more precisely, the estimates $V_1\{4\}^4$ and 
$V_2\{4\}^4$) as functions of $c_1\{4\}$ and $c_2\{4\}$. 

Let us now consider differential flow. 
Integrated and differential flows may be measured using two different detectors: 
for instance, a large acceptance detector for integrated flow, and a smaller 
one, but with better particle identification or $p_T$ determination, for 
differential flow. 
For sake of generality, we thus denote by $A'(j,\psi,p_T,y)$ the corresponding 
acceptance function and by $A'_k(j,p_T,y)$ its Fourier coefficients defined as 
in (\ref{ak_pTy}). 
The differential acceptance coefficients $a'_k$ are then defined as in 
Eq.\ (\ref{ak}), without the weights and the summation over $j$ (since one 
usually measures the differential flow of identified particles) and with the 
integration over $p_T$ and $y$ restricted to the phase-space region under 
interest (typically, one integrates over $p_T$ {\em or} $y$, so as to obtain 
$v_n$ as a function of $y$ or $p_T$, respectively).    

Once again, anisotropies in the detector acceptance lead to some interference 
between the flow harmonics in the expressions of the cumulants 
$d_{mn/n}\{2k+m+1\}$. 
Here are, for example, the relations between the lowest order cumulants and 
flow which allow one to extract the differential directed flow $v'_1$ and the 
differential elliptic flow $v'_2$ obtained either with respect to $V_1$ or 
$V_2$, i.e., what we denote $v'_{2/1}$ and $v'_{2/2}$ (relations for other 
cumulants can be found in Ref.\ \cite{Borghini:2001vi}, Appendix C2): 
\begin{mathletters}
\label{diff_acc}
\begin{eqnarray}
\label{d11-4}
d_{1/1}\{4\}&=& -{\rm Re}\left[ (1-|a_1|^2)\left((1-|a_1|^2)^2 +
2 \left|a_2-a_1^2\right|^2\right)
+(a'_2)^* (a_2-a_1^2)\left(2(1-|a_1|^2)^2 +
\left|a_2-a_1^2\right|^2\right)\right] v'_1 V_1^3\cr
& & - 
{\rm Re}\left[ a'_1 (a_1^*-a_2^* a_1)\left( \left|a_1^*-a_2^* a_1\right|^2+
2\left|a_3-a_1a_2\right|^2\right) \right. \cr
& & \qquad \left.
+ (a'_3)^* (a_3-a_1a_2)\left(2\left|a_1^*-a_2^* a_1\right|^2+
\left|a_3-a_1a_2\right|^2\right)\right] v'_2 V_2^3 , \\
d_{2/1}\{3\} & = &
{\rm Re}\left[ (1-|a_1|^2)^2+(a'_4)^* (a_2-a_1^2)^2\right] v'_2 V_1^2 \cr
&  & +\ 2\,{\rm Re}\left[ (a'_1)^* (a_1^*-a_2^* a_1) (a_2-a_1^2) + 
(a'_3)^* (a_3-a_2a_1)(1-|a_1|^2)\right] \,v'_1 V_1 V_2, \label{d21-3}\\
\label{d22-4}
d_{2/2}\{4\}&=& 
-{\rm Re}\left[ (1-|a_2|^2)\left((1-|a_2|^2)^2 +
2 \left|a_4-a_2^2\right|^2\right)
+(a'_4)^* (a_4-a_2^2)\left(2(1-|a_2|^2)^2 +
\left|a_4-a_2^2\right|^2\right)\right] v'_2  V_2^3\cr
& &  -
{\rm Re}\left[ (a'_1)^* (a_1-a_2 a_1^*)\left( \left|a_1^*-a_2^* a_1\right|^2+
2\left|a_3-a_1a_2\right|^2\right) \right. \cr
 & & \qquad \left. 
+(a'_3)^* (a_3-a_1a_2)\left(2\left|a_1^*-a_2^* a_1\right|^2+
\left|a_3-a_1a_2\right|^2\right)\right] v'_1  V_1^3. 
\end{eqnarray}
\end{mathletters}
In these expressions, we have neglected terms involving higher flow harmonics 
$V_{n\geq 3}$ or $v'_{n\geq 3}$, and ${\rm Re}$ means that one should take the 
real part. 
Of course, they reduce to Eqs.\ (\ref{d_and_v'}) when the acceptance is 
isotropic. 

Any two of Eqs.\ (\ref{diff_acc}) constitute a linear system which can be 
inverted to obtain $v'_1$ and $v'_2$ once $V_1$ and $V_2$ have been extracted.

\section{Additional remarks and comments}

The cumulant method not only eliminates nonflow correlations, it also provides 
several independent estimates of the flow from cumulants of various orders. 
Since nonflow correlations between 4 or more particles are expected to be 
negligible, $V_n\{4\}$ and $V_n\{6\}$ should be consistent with each other 
within statistical error bars. 
With the high multiplicity produced at ultrarelativistic energies and the large 
acceptance detectors available, estimates show that one could even construct 
cumulants of 8-, 10-particle correlations.
Since the generating function formalism yields all cumulant orders at once, one 
would simply need to increase the number of interpolation points $z_{p,q}$, 
which would result in a moderate increase in computer time. 
Checking that all higher order cumulants yield compatible values of $V_n$ would 
give the first direct evidence that azimuthal correlations are of collective origin. 
Similarly, in the case of differential flow, various cumulants orders yield 
independent estimates which can be compared to one another. 

Other consistency checks can be proposed to test the reliability of the 
results. 
First, one should perform the analysis with at least two values of the 
parameter $r_0$ entering Eq.\ (\ref{points}), in order to check the stability 
against numerical errors. 
Then, one can try different values of $M$ in Eq.\ (\ref{gen_func}), as, for 
instance, $M=0.8\mean{M_{\rm tot}}$ and $M=0.6\mean{M_{\rm tot}}$. 
That may also be a way to increase the statistics. 
In the same spirit, it is still possible to let $M$ vary from event to event in 
Eq.\ (\ref{gen_func}), taking $M=M_{\rm tot}$, especially if the acceptance is 
reasonably good. 
In that case, $M$ must be replaced by the average $\mean{M_{\rm tot}}$ in 
Eq.\ (\ref{int_cum}), and one should check that the results are consistent with 
what would be obtained with a fixed $M$. 

The last consistency check regards differential flow. 
We have seen that the integrated flow $V_n$ can be calculated using different 
weights $w_j$; the resulting weighted averages $V_n$'s differ. 
However, the values of the differential flow $v'_{mn/n}\{2k+m+1\}$ 
obtained with different weights should be consistent within error
bars if they are not contaminated by nonflow effects.

The recent STAR analysis \cite{Tang:2001yq} illustrates quite well the 
relevance of cumulants to the flow analysis: for integrated flow, the lowest 
order $V_n\{2\}$ reproduces the results of the standard two-particle methods, 
as it should. 
On the other hand, the higher order estimate $V_n\{4\}$ is lower than $V_n\{2\}$ 
beyond statistical error bars, especially for peripheral collisions, thereby 
suggesting that nonflow correlations are important. 
Although statistical uncertainties on higher order cumulants are larger, this 
loss is therefore compensated by the gain on errors from nonflow effects.

\section*{Acknowledgments}

We thank N.\ N.\ Ajitanand, M.\ Ga\'zdzicki, R.\ Lacey, J.\ Milo\v sevi\'c, 
A.\ M.\ Poskanzer, H.\ Str\"obele, A.\ Tang and A.\ Wetzler for helpful 
suggestions. 
N.\ B.\ acknowledges support from the ``Actions de Recherche Concert\'ees'' 
of ``Communaut\'e Fran\c{c}aise de Belgique'' and IISN-Belgium.

\end{document}